XString: XML as a String
William Gilreath
will@williamgilreath.com
wgilreath@gmail.com

October 2006

Abstract

Extensible markup language (XML) is a technology that has been much hyped, so
that XML has become an industry buzzword. Behind the hype is a powerful
technology for data representation in a platform independent manner. As a
text document, however, XML suffers from being too bloated, and requires an
XML parser to access and manipulate it. XString is an encoding method for
XML, in essence, a markup language's markup language. XString gives the
benefit of compressing XML, and allows for easy manipulation and processing
of XML source as a very long string.

Keywords: document processing, document encoding, encoding, extensible
markup, markup language, XML

1. Overview of XML

XML is a way of marking up data with custom tags in order to describe it. XML provides for marking up data, but also organizing the data hierarchically, providing a structure of the data as well as an associated description. To insure uniform structure, XML requires all documents, regardless of the tags created, to follow rules of well-formedness:

1. Each document must possess a unique root element.
2. Open tags must have close tags (pair-wise matching of tags).
3. The version of XML, and encoding must be at the start of a document.
4. Tags can be nested, but not overlapped within the document.
5. Attributes with names and optional values can only be used for open tags.
6. Attributes with a value can only have a single value within quotes.
7. Elements for specific tags must follow certain naming conventions.
8. Certain predefined entities may be used instead of character [XML].

Beyond well-formedness an XML document can be verified against its structural specification, which is a document type declaration (DTD) or XSchema. Hence an XML document does not need to be valid or validated, but it must be well-formed.

With the ability to define custom markup tags, and a general way to describe document structure that follows certain rules of formation, XML has enormous potential as a generalized means for data markup and description for a domain specific knowledge.

1.2 Problems Associated with XML

While XML offers enormous potential in terms of domain-specific data markup and description, there are two major problems with XML. The problems include:

1. Data bloating and verbosity from the markup of data,
2. Complexity for access and manipulation of XML.

Data bloating occurs from the nesting and structuring of tags. In large data sets, markup can assist in document manipulation, but for small data sets it can bloat the corresponding data.

For example, a simple list of properties in an initialization file (such as the now obsolete Window INI files) would be:

```
[ENVIRONMENT]
TERM=ANSI
CURRENCY=DOLLAR
KEYBOARD=PC101
HOME=/home/jdoe
USER=jdoe
```

This is a fairly straightforward organization of the information, and easily processed to get the required information for each property. Expressed in XML markup, the initialization file would be:

```
<ENVIRONMENT>
<TERM>ANSI</TERM>
<CURRENCY>DOLLAR</CURRENCY>
<KEYBOARD>PC101</KEYBOARD>
<USER>jdoe</USER>
</ENVIRONMENT>
```

The XML markup for the properties has effectively doubled the descriptive information from the duplication of tags. For smaller data sets, XML markup can lead to data increase, making XML unwieldy and space inefficient.

Access and manipulation of an XML document is accomplished with an XML parser. The predominant XML parser is for the document object model (DOM), and is very large and complex software. The DOM parser specification from the World Wide Web Consortium (W3C) is well over 100-pages. The complexity involved in manipulating a simple XML document requires an equally complex XML parser. Moreover, there is no simple way to manipulate XML without first parsing. An alternative XML parser is the simple API for XML (SAX), but SAX is an event-based parser as opposed to DOM, which is tree-based.

1.3 Other XML Technologies

XML is not just a technology of extensible markup for domain-specific documents. The extensible markup is one facet of an entire group of technologies which are based on the basic principles of XML. It seems plausible that one of the other XML technologies could serve as a solution to the problem of XML bloating and difficulty in handling. However, the other XML technologies are not suitable, although the two most likely to work as a solution are XPath and XLink.

1.3.1 Xpath

XPath has a similar syntax to XString [Xpath], so it would appear that Xpath can be used instead of XString. XPath allows for a general expression as a path through the XML document as a hierarchical tree structure. XPath can be used as a query to return nodes that are within a specified path, or perform pattern matching; such as in XML Style sheet Language Transformations (XSLT).

However, XPath does not represent an entire document, rather XPath is a way to address only a part of the document.

1.3.2 XLink

XLink is conceptually similar to Xpath [Xlink], however XLink allows XML documents to link to other XML documents and other resources. In effect, XLink address an external document and creates a hyperlink to it. XLink uses XPointers to locate resources, and an XPointer is an extension to XPath but better suited for linking. A primary difference between the XPointer and XPath is that XPath does not specify context, it is evaluated within a context. XPointer does specify a context.

1.3.3 Other XML Technologies

There are other XML technologies such as XSLT, the XML Expression language
(XEXPR), the XML Query Language (XQL), and so forth. However each technology
is intended to provide a solution to a particular problem, such as XSLT for
document transformation, XQL for data extraction, XEXPR for script-like
functionality, and so forth. Yet there is no existing technology to supply a
solution to the problem addressed by XString, focusing on the problem of
concise and compact XML representation.

1.4 Other Compression Schemes

There is little work on the compression of XML. Liefke and Suciu describe,
XMill, an XML compression algorithm that transforms documents to expose
redundancy then applies standard text compressors, such as GZIP to it
[Liefke]. However, a major drawback of XMill is that it can not be used for
online processing of compressed documents and it actually hinders compressors
other than gzip, and it requires user assistance to achieve the best
compression [Cheyney].

Cheyney describes two schemes for hierarchical compression using online
binary encoding. But these use existing commercial compressors such as GZIP,
and have significant overhead processing time [Cheyney].
Cannataro et al describe a compression scheme for XML called SqueezeX, which
is based on "a multidimensional/classification-based interpretation of schema
ad on the application of semantic and type-oriented compression" [Cannataro].

However, the approach is a lossless scheme. Finally, another related
technology is the PYX notation for XML [Pyxie] The acronym PyX stands for
PostScript + Python + TeX. PyX is a line-oriented notation; -- each line of
PYX contains the information for a single parsing event as an alternative to
the DOM and SAX XML parsers. But PyX is intended as an alternative means for
parsing and processing XML, not for compression.

2. XString

Our research and experimentation have led us to develop a text based encoding
system for XML that is fast, has high compression rates, and is suitable for
online processing.

XString is a concise description of XML, in essence XML's XML. XString
eliminates the two problems of XML, data bloat and complexity of data
manipulation. XML documents can be converted to and from XString, and
conversion to XString can compress the size of an XML document, yet still be
able to manipulate and access the data. XString uses a similar syntax to
XPath, to retain a familiar syntax.

2.1 From XML to XString

XString is a unique prefixing code, that is, there is a unique character that
prefixes a tag element to identify it and the tag type. The unique prefixing
scheme eliminates the need for a suffix, or close tag, which eliminates the
extra bloat XML adds to data.

To illustrate, the previous properties file can be written using LISP, which is based upon Cambridge Polish (prefix) notation, as opposed to Reverse
Polish (postfix) notation. The LISP-like markup would then be:

```
(ENVIRONMENT
(TERM ANSI)
(CURRENCY DOLLAR)
(KEYBOARD PC101)
(USER jdoe)
)
```

The closing </...> tag has been eliminated, yet the document structure and
data markup is preserved.

One problem in this scheme is immediately observed which tag is which? The
second element is unknown in terms of what kind of tag it is. Another
observation is that the closing parenthesis occurs in a regular pattern.
Eliminating the closing tag, but replacing it with a single character is not
completely efficient.

The solution is to use a unique prefix character which identifies the tag
element kind, and its use. The solution to the redundant closing parenthesis
is to use each tag as a prefix to others, with implicit closing of the tags.

Taking the LISP-like example and making it a flat string:

```
(ENVIRONMENT(TERM ANSI)(CURRENCY DOLLAR)(KEYBOARD PC101)(USER jdoe))
```

Now as a long string, the redundancy of the closing parenthesis is
immediately apparent, along with that each starting tag is a prefix, but also
acts as an implicit suffix to close the previous tags. Rewriting the string
yields:

```
(ENVIRONMENT(TERM ANSI(CURRENCY DOLLAR(KEYBOARD PC101(USER jdoe
```

## 2.2 XString Prefix Characters

The tagging resolves the problem of the redundant parenthesis, but the
problem of what kind of tag remains. In this illustrative example, there are
two tag types, a general markup tag, and a text tag which is data. This
example has defaulted to the open parenthesis ( and a white space as the
prefix characters. Using the XString prefix characters it becomes:

```
/ENVIRONMENT/TERM'ANSI|CURRENCY'DOLLAR|KEYBOARD'PC101|USER'jdoe
```

By method of illustration, each prefix character effectively suffixes (or
closes) each tag, and also identifies the tag element kind.

Document Type Definition (DTD) elements (prefixed with an exclamation mark! )
are from XML's predecessor, Standardized Generalized Markup Language (SGML)
[SGML] which used a DTD to define a document structure. DTD's are still
widely used in defining XML documents, although the use of XML schemas
(XSchemas) to define XML documents in XML is a new and more powerful than
DTD's. However, because DTD elements can appear in XML documents, the ability
to encode the DTD elements is included as part of XString.

For example a DTD element in a HTML document might be:

```
<!DOCTYPE html PUBLIC "-//W3C//DTD HTML 4.01 Transitional//EN">
<html>
...
</html>
```

One other possible tag or element to be represented in XString is an entity element. Entity elements represent characters in XML, often times the special characters used in XML. For example < is the entity for a less-than character or <, but is a named entity. Other characters are represented by their numeric entity, such as < which is again the less-than sign. Entities are not explicitly delineated as another type of node. Often, an entity will be embedded within a text node. Entities are effectively special encoding of a text character. No reason is needed to distinguish the entity from the rest of the text as a special kind of node.

Table 1 summarizes each prefix character for each tag, and gives an example of the XML and corresponding XString. The XString prefix characters try to follow the characters used in XPath, although other characters for prefixes are unique to XString.

| Tag | Prefix | XML | XString |
|------|--------|-----|---------|
| Child Tag | / | <X><Y></X> | /X/Y |
| Sibling Tag | \| | <X><Y><Z></X> | /X/Y\|Z |
| Comment | @ | <!-- comment --> | -comment |
| Processing Instruction | = | <?SQL select * from all?> | ?SQL select * from all |
| CData | [ | <!CDATA[0xFFFF]]> | [0xFFFF |
| DTD Element | ! | <!DOCTYPE html> | !DOCTYPE html |
| Text | ` | <X>text<X> | /X'text |
| Text dual | " | <X>huh?</X> | /X"huh?" |
| Attribute Name | – | <X NAME/> | /X@NAME |
| Attribute Value | ? | <X NAME='Jon' /> | /X@NAME=Jon |

Table 1: Unique XString Prefix Characters

Two variants exist for a regular tag <...> and for a text. The variant for the start tag is for a child (nested) tag, and for a sibling tag. The distinction is to indicate nesting of tags as opposed to tags which are not nested but contained within. The two variants for the text tag allow for a text tag to be explicitly delineated, one with the regular prefix tag of a single quote, the other with a dual starting and closing double quotes. The double quotes allow for enclosing characters which could be confused for other characters, such as a question mark.

XString being a markup of XML, follows the same rules of well formedness for XML. The distinction between child tags which are nested, and tags which are siblings are needed to follow the rules of well-formedness.

## 2.3 XString Ambiguity in representing XML Document

While the scheme described above will be shown to lead to good compression, there are some problems introduced by the XString algorithm, however.

### 2.3.1 Structural Depth

While XString is an efficient encoding and representation of XML, there is one problem: ambiguity in nesting tags. Consider the two XML fragments:

```
<XML> <XML>
<TAG> <TAG>
<?PI?> <?PI?>
<!--comment--> </TAG>
text <!--comment-->
</TAG> text
</XML> </XML>
```

Both variants have the same XString representation:

```
/XML/TAG?PI-comment'text
```

This ambiguous nesting occurs because the XML is not nesting tags which are sibling or child, but the non-data tags of processing instruction, comment, and text or unary tags. There is no clearly explicit way to know where the closing tag of the child or sibling is positioned. The ambiguity of nesting tags is somewhat analogous to the "dangling else" problem in Pascal. [MacLennan p. 147]

The simple solution is to have a meta-prefix of how deep or the depth of the starting tag. The depth is indicated with a plus + followed by an integer value. Rewriting the two XML examples appropriately:

```
<XML> <XML>
<TAG> <TAG>
<?PI?> <?PI?>
<!--comment--> </TAG>
text <!--comment-->
</TAG> text
</XML> </XML>
```

```
/XML/TAG+3?PI-comment'text /XML/TAG+1-comment'text
```

The prefix of a plus indicates a depth marker, and the meta-prefix is the depth of the preceding start tag. Now there is no ambiguity and each variant of the XML has a clearly unique XString representation.

### 2.3.2 Sibling and Child Node Dependencies

In an XML document all nodes are children of the root node. This follows one of the principles of well-formed XML, that all nodes have a unique parent node. A parent node contains child nodes, and the root node of an XML document is the highest parent node.

A sibling node is a node that shares a common parent node with other nodes that are children of the same parent. Sibling nodes are the children of a parent node, but are siblings to each other. A sibling node is at the same level of an XML document as other siblings.

Essentially, nodes contained within a node are children to a parent node, but the contained nodes are sibling nodes with respect to one another. The type of node, sibling or child, is dependent upon relating to a parent node, or a peer node.

The simple XML document:

```
<ROOT>
<CHILD1>john doe<?PI?></CHILD1>
<CHILD2>jane doh</CHILD2>
</ROOT>
```

The three child nodes CHILD1, CHILD2, CHILD3 are all children with respect to the parent node, ROOT. However, the nodes are also siblings in respect to each other.

The XString representation is:

```
/ROOT/CHILD1 john doe?PI|CHILD2 jane doe
```

in the XString form, only the first node CHILD1 is prefixed as a child node. The remaining node is prefixed as a sibling node, which indicates the next node shares the same parent as the previous node.
The sibling node is an implicit end of depth marker from the previous child node. Essentially the previous node extends and encompasses to the sibling node.

```
/ROOT/CHILD1 john doe?PI|CHILD2 jane doe
^ depth to sibling ^
```

The sibling prefix is an implicit depth suffix within the structure of the XML document. Rewriting the XString using depth prefix explicitly:

```
/ROOT/CHILD1+2 john doe?PI/CHILD2 jane doe
```

The sibling prefix implicitly delineates the depth of a previous child node, and indicates that the next node is a sibling, or shares the same parent node.

2.3.3 Depth Nodes conflicting with Sibling Nodes

The sibling node is an implicit depth indicator, and a depth prefix is explicit depth indicator. It is possible that a depth prefix can conflict with the implicit depth from a sibling node. A conflict in the depth of one node with a sibling node can lead to ambiguous and ill-formed XML, mangling the original document structure.

For example the XML document:

```
<ROOT>
<CHILD/>
<!-- comment -->
<![CDATA[0xFF]]>
<SIBLING/>
</ROOT>
```

The valid (maintains XML document structure) XString is:

```
/ROOT/CHILD+0-comment[0xFF/SIBLING
```

without the depth prefix, the XString is:

```
/ROOT/CHILD-comment[0xFF|SIBLING
```

Unfortunately, the XString has altered the XML document structure; when reconstructed from the XString, the XML is:

```
<ROOT>
<CHILD>
<!-- comment -->
<![CDATA[0xFF]]>
</CHILD>
<SIBLING/>
</ROOT>
```

Hence the depth of the node CHILD needs to be clearly indicated explicitly. The initial attempt takes the original XString and inserts the depth prefix:

```
/ROOT/CHILD+0-comment[0xFF|SIBLING
```

The depth prefix of +0 indicates CHILD does not enclose or is a parent to the comment and CDATA nodes. Unfortunately, this rewrite of the XString mangles the XML, when reconstructed:

```
<ROOT>
<CHILD>
<!-- comment -->
<![CDATA[0xFF]]>
</CHILD>
</ROOT>
<SIBLING/>
```

Not only is the XML document structure altered from the original, it is worse in that the resulting XML is not well formed.

```
/ROOT/CHILD+0-comment[0xFF|SIBLING
^ implict depth ^
```

The fatal flaw is that the sibling node SIBLING has created an implicit depth back to the CHILD node, and this conflicts with the explicit depth indicated. To correct the XString, and still retain the integrity of the structure of the original XML document, the sibling node is changed into a child node.
The XString is:

    /ROOT/CHILD+0-comment[0xFF/SIBLING

which is the original XString given as valid for the XML document.

The problem is that conflicts between explicit depth and implicit depth of
sibling nodes can contradict each other, leading to mangled XML.

2.3.4 Solving the Depth Problems with Prefix Coding

One straightforward solution to the problem of conflict with depth tags and
sibling tags is to represent an XML document purely as a document with only
children and depth tags. The child-depth prefix coding would have to supply
implicit depth that sibling tags had previously provided. So the encoding can
be less efficient, but there is no ambiguity of conflict between sibling
nodes and depth nodes.

The process of converting a XString with sibling node elements with implicit
depth to an explicit depth with child nodes is straightforward.
For example, given the XString:

    /EMP/ROW/FNAME'John|LNAME'Doh|ROW/FNAME'Jane|LNAME'Doh

The process of mapping the sibling nodes operates on the XString left to
right. For each sibling, it is converted to a child, with the depth for the
following tags to the next sibling or the end of the XString. The previous
child tag needs the depth to the sibling explicitly

    /EMP/ROW/FNAME'John|LNAME'Doh|ROW/FNAME'Jane|LNAME'Doh

The first child tag has a text tag, and then is followed by the sibling node
LNAME. The child tag does not need to be changed, but the depth needs to be
explicit.

    /EMP/ROW/FNAME+1'John|LNAME'Doh|ROW/FNAME'Jane|LNAME'Doh
    /EMP/ROW/FNAME'John|LNAME'Doh|ROW/FNAME'Jane|LNAME'Doh

Similarly, the LNAME tag encloses the text tag 'Doh.

    /EMP/ROW/FNAME+1'John/LNAME+1'Doh|ROW/FNAME'Jane|LNAME'Doh

The ROW tag is a sibling, and more importantly, is at the same depth as the
previous ROW tag, which is a child of the root node EMP. The second ROW tag
is likewise a child of the root node EMP.

    /EMP/ROW/FNAME+1'John/LNAME+1'Doh|ROW/FNAME'Jane|LNAME'Doh

Instead of the sibling ROW referring back to the prior sibling node, it
refers effectively back to the first ROW tag. As a sibling it is at the same
depth, and a child of the root node /EMP.

By referring back to the first ROW tag, it has shown that the prior node which it implicitly is indicating the depth of is not explicit enough. Both ROW tags as children need the depth to be explicit, so that the nodes have an explicit parent node.

The first ROW tag encloses 4 tags, the depth tag is not counted among them. So the revamped XString becomes:

    /EMP/ROW+4/FNAME+1'John/LNAME+1'Doh|ROW/FNAME'Jane|LNAME'Doh

The child ROW tag now is explicitly enclosing the 4 child nodes, up to the next ROW tag, which is a sibling. The sibling ROW tag now refers to the prior ROW tag, which has its depth clearly indicated.

The XString with the sibling ROW:

    /EMP/ROW+4/FNAME+1'John/LNAME+1'Doh|ROW/FNAME'Jane|LNAME'Doh

The sibling ROW can be a child node, and it encloses the next 4 tags. So the XString is then:

    /EMP/ROW+4/FNAME+1'John/LNAME+1'Doh/ROW+4/FNAME'Jane|LNAME'Doh

The FNAME tag is a child node, but because the next tag is sibling, it must have the depth explicitly delineated. So the XString becomes:

    /EMP/ROW+4/FNAME+1'John/LNAME+1'Doh/ROW+4/FNAME+1'Jane|LNAME'Doh

The last tag is sibling LNAME, and it is converted to a child tag, and has the depth explicitly marked as:

    /EMP/ROW+4/FNAME+1'John/LNAME+1'Doh/ROW+4/FNAME+1'Jane|LNAME'Doh

the XString then becomes:

    /EMP/ROW+4/FNAME+1'John/LNAME+1'Doh/ROW+4/FNAME+1'Jane/LNAME+1'Doh

The process of converting a sibling XString to a child-depth XString is finished.

Comparing two XStrings:

Original: /EMP/ROW/FNAME'John|LNAME'Doh|ROW/FNAME'Jane|LNAME'Doh

Revised: /EMP/ROW+4/FNAME+1'John/LNAME+1'Doh/ROW+4/FNAME+1'Jane/LNAME+1'Doh

The revised XString has grown in size that is expected as the implicit information in the siblings as in regard to depth is now explicit. However the structure of the XML is intact, and the revised XString is a compact representation of XML.

Likewise, the only node not changed that is not a text node is the root tag /EMP. It is clearly the parent of all the nodes, so its depth would enclose all the following nodes, for a depth of 10. (Add the two ROW depth, and then include the ROW tags in the count 4 + 4 + 2 = 10.)

2.3.5 Encoding of Prefix Characters used as Data

A complexity with using XString is that the prefix characters can occur as part of the data in a comment, CDATA node, or text node. Distinguishing the XString prefix characters from the regular characters in the XML document is needed.

One solution is to use entity elements as the encoding for the special XString prefix character. The entity element is a literal character, and is distinct from the XString prefix character set. As an example:

        <GREETING>'Hello World!!!'</GREETING>

compresses in Xstring to

        /GREETING''Hello World!!!'
The single quote ' is ambiguous. But, using entities for the right and let quotes rather than literals, yields

        /GREETING'‘Hello World!!!’ .

Another approach to delineating the XString prefix characters is to use a sentinel character as a prefix to the prefix characters. For example, the null character (ASCII/Unicode value 0) could be used to prefix each XString prefix character. So then the XML:

        <GREETING>'Hello World!!!'</GREETING>

in XString with a null character:

        \0/GREETING\0''Hello World!!!'

Hence after the null character the next character clearly is the XString prefix. The disadvantage is that the null character makes each XString prefix two characters in size, instead of one.

For either approach, using entity encoding, or a sentinel character, the use limited to distinguish the XString prefix characters would reduce the cost of using either alternative.

2.4 XString Efficiency

XString compresses XML by more efficiently encoding it, removing redundancy. It is of interest to know how much compression is possible, and in what kinds of XML the most efficient compression will result.

Table 2 summarizes the character size for each XML element and the corresponding XString character size. Overall for XString, the size is often n+1 characters, n being the information in the specific tag. Attributes, the name and value are each n+1, so overall they are m+n+2. The double-quoted text is the only XString expression, which has n+2.

| XML Elements | XML Regular Expression | XML Size | String Regular Expression | XString Size |
|---|---|---|---|---|
| Nested tags | <(char)+>...</(char)+> | 2n+5 | \(char)+... | n+1 |
| Empty tag | <(char)+/> | n+3 | \(char)+ | n+1 |
| PI tag | <?(char)+?> | n+4 | ?(char)+ | n+1 |
| DTD element | <!(char)+> | n+3 | !(char)+ | n+1 |
| Comment tag | <!--(char)+--> | n+7 | -(char)+ | n+1 |
| CDATA tag | <![CDATA[(char)+]]> | N+12 | [(char)+ | n+1 |
| Text tag | (char)+ | n | '(char)+ | n+1 |
| Text tag | (char)+ | n | "(char)+" | n+2 |
| Attributes | <...(char)+= (char)+ > | m+n+3 | \...@(char)+=(char)+ | m+n+2 |

Table 2: Summarizing XML to XString Efficiency

Overall, there is some reduction in size for most XML elements. The exception is text tags, which can actually increase by one or two characters. The largest increase is for nested and start/end tags, where the size of 2n+5 characters are reduced to n+1. In terms of efficiency, at the base size of n=1 is 2/7. Using calculus and taking the limit of n the ratio is 1/2. Hence highly nested tags will result in the greatest efficiency in XString.

Conversely, lots of text will hinder XString efficiency. Other XML elements will provide some reduction in characters, but not as great as nested elements.

For XML nesting with ambiguity that requires the depth prefix, some characters are added into the XString representation. With a high degree of nesting, the additional overhead to delineate depth can be minor. For example, there is a version of XML called Simple XML that uses nesting not attributes [Simple]. This makes it ideal for use with XString. Of course, ColdFusion Markup Language is similarly amenable to XString [Weiss]

However, when the nesting level is low there is only a few characters saved.

For example

        <XML TYPE="1.0">

converts to

        /XML@TYPE=1.0

but the type gains one prefix character. The attribute value type looses the double quotes, but it is a marginal gain. The best gain is on nested tags because you have 2n+5 characters reduced to n+1.

The main feature that has made XML so appealing, the nesting and double tags to markup data, is the exact feature that can allow for substantial reduction by conversion to XString. Given that an XML document will have nested elements and markup, along with other tags and text, the reduction is idealistically at 50% of the original XML document.

This can be determined with elementary calculus with the limit as n approaches infinity, and L`Hopital's rule for indeterminate ratios:

$$\lim_{n \to \infty} \frac{n+1}{2 \cdot n + 2} = \frac{d}{dn} \lim_{n \to \infty} \frac{n+1}{2 \cdot n + 2} = \lim_{n \to \infty} \frac{1}{2} = \frac{1}{2}$$

Yielding a 50% compression ratio.

2.5 Example of XML to XString

An example of converting XML to XString a simple database table with FIRSTNAME and LASTNAME are represented in two forms as XML. One uses nested markup of the rows, the other attributes.

Row-oriented mapping XML would require 142 characters and look like the following:

```
<EMP>
<ROW>
<FNAME>John</FNAME>
<LNAME>Doe</LNAME>
</ROW>
<ROW>
<FNAME>Jane</FNAME>
<LNAME>Doh</LNAME>
</ROW>
</EMP>

/EMP/ROW/FNAME'John|LNAME'Doh|ROW/FNAME'Jane|LNAME'Doh (54 characters)
```

Using the nested markup, the row-oriented XML document of 142 characters is reduced to a XString the length of 55 characters, saving 87 characters for a reduction of 61%.

Column-oriented mapping XML, requiring 54 characters, would appear as

```
<EMP>
<COL>
<FNAME>John</FNAME>
<FNAME>Jane</FNAME>
</COL>
<COL>
<LNAME>Doe</LNAME>
<LNAME>Doh</LNAME>
</COL>
</EMP>
```

/EMP/COL/FNAME'John│FNAME'Jane│COL/LNAME'Doe│LNAME'Doh

Record-oriented mapping XML might require 55 characters as the following:

```
<EMP>
<REC FNAME="John" LNAME="Doe"/>
<REC FNAME="Jane" LNAME="Doh"/>
</EMP>
```

(55 characters)
/EMP/REC@FNAME=John@LNAME=Doe│REC@FNAME=Jane@LNAME=Doh (83 characters)

The record oriented XML document has the same size of XString, 55 characters.
But the original document was of size 83 characters, for a reduction of 28
characters saving 34%.

It is interesting to note that the same XString in terms of length comes from
two very different XML documents of different sizes. Interestingly as well, a
side-by-side comparison shows the same information, just organized
differently within the XString encodings of the XML.

/EMP/ROW/FNAME'John│LNAME'Doh│ROW/FNAME'Jane│LNAME'Doh

/EMP/REC@FNAME=John@LNAME=Doe│REC@FNAME=Jane@LNAME=Doh

The other problem with XML, the difficulty in manipulation, is not a problem
for XString. Using basic string operations and not an XML parser, the XString
representation for an XML document with attributes could be converted to an
XString document without attributes. <example with C string operators?>
The substitution is for the attribute value prefix of = to be converted to a
text '. Then the attribute name prefix of @ is converted to a child \ or
sibling node prefix, depending upon previous tags. With this direct
substitution, one XML document is transformed into another. XString would
make
an excellent intermediary form to manipulate, access, and search an XML
document.

As a final comparison, the original properties file, with XML, and XString (using carriage returns and new lines) representation are given:

```
[ENVIRONMENT]        <ENVIRONMENT>                        /ENVIRONMENT
TERM=ANSI            <TERM>ANSI</TERM>                    /TERM'ANSI
CURRENCY=DOLLAR      <CURRENCY>DOLLAR</CURRENCY>          |CURRENCY'DOLLAR
KEYBOARD=PC101       <KEYBOARD>PC101</KEYBOARD>           |KEYBOARD'PC101
USER=jdoe            <USER>jdoe</USER>                    |USER'jdoe
                     </ENVIRONMENT>
```

With some reformatting to compare line for line, the XML is bloated with markup tags on each item from the properties list. The XString only adds one character per line, and corresponds directly to the properties listing. Hence XString gets the benefit of XML, without adding unnecessary tags to the data.

## 2.6 Alternative Encoding of XString

XString is simply encoded as text, much like XML is encoded in a text document. But there are other means for encoding XString, with an improvement in the length of the XString so encoded. The added compression of the XString, along with it still being in a form to process make alternative encodings useful, depending upon the nature of the application or use.

### 2.6.1 Redundant substitution

A possible representation of XString is to remove redundant long text tags and nodes, and use a numeric key to substitute. This approach requires another unique prefix character, such as the hash or pound sign #. Then wherever the substituted tag name occurs, the numeric value is used instead. A simple example would be:

/XML/AVERYLONGTAGNAME'child1|AVERYLONGTAGNAME'child2

the node or tag AVERYLONGTAGNAME is redundant and a long string. So with a redundant substitution:

/XML/AVERYLONGTAGNAME#0'child1|0'child2

the long string is factored out and replaced with the integer value of 0. One important restriction would be that no tag names or nodes that are to be replaced are integers.

### 2.6.2 Byte/binary representation

Another alternative encoding rather than complete Unicode/US-ASCII (or other text encoding schemes) for the prefix characters, is to encode the prefix characters as bytes. With ten unique prefix characters, each byte can encode two prefix characters (each of the 4-bits can represent 16 unique prefix characters). The depth prefix can be included in the list of unique prefix characters. For each string, which is prefixed, the length, followed by the actual encoded characters can be used.

For example an XString element could be encoded as:

Xelement ::= <prefix byte><string length><character string><string
length><character string>

/XML'XMLTEXT in Unicode 2 bytes * 12 characters = 24 bytes, or in UTF-8 12
bytes as regular text:

UTF-8: 2F 58 4D 4C 27 54 45 58 54
Unicode: FF FE 2F 00 58 00 4D 00 4C 00 27 00 54 00 45 00 58 00 54 00

As the prefix encoding:

UTF-8: OF|03|58 4D 4C|07|58 4D 4C 54 45 58 54
Unicode: OF|03|00 58 00 4D 00 4C|07|00 58 00 4D 00 4C 00 54 00 45 00 58 00 54

OF indicates child followed by text
03 indicates 3-bytes for first string
58 = X
4D = M
4C = L
07 indicates 7-bytes for next string
54 = T
45 = E
58 = X
54 = T
58 = X
4D = M
4C = L

With a longer XString with recurrent prefixes, the encoding can reduce 2-
bytes for UTF-8 characters to 1-byte, and 4-bytes of Unicode to 2-bytes.
Other encoding systems such as using the unique prefix along with null 0x00
as a separator. The prior string then is:

As the prefix encoding:

UTF-8: OF | 58 4D 4C | 00 | 58 4D 4C 54 45 58 54
Unicode: OF | 00 58 00 4D 00 4C | 00 | 00 58 00 4D 00 4C 00 54 00 45 00 58 00
54

If a unique prefix is used, it is the separator; otherwise the null character
is used. The encoding reduces one byte, so UTF-8 is 12-bytes (unchanged), and
the Unicode is 22 bytes. A longer XString would lead to more redundancy.

Another possible encoding system is to use a table of strings, and then
substitute the string with the index into the table, so that very long and
repetitive strings are more concisely encoded. In essence, XString serves as
the template for the encoding, and the prefix characters can be compactly
encode as bytes. A byte or binary encoding of a XString can more concisely
encode an XString, the disadvantage that the XString as a binary
representation becomes quasi-proprietary in the format.

## 3. Applications of XString

### 3.1 Compact Data Storage

One of the best uses of XString is in the compact representation of an XML document. As has been shown, XString compresses an XML document more effectively with a more concise encoding. A more compact representation takes less storage space, and in sending and receiving information across a network, takes less time to transfer information. With the rise of web services, and ubiquity of the World Wide Web, savings in time and storage are highly significant.

### 3.2 Web Server with XHTML

The impact of XML has had reverberations beyond the much touted areas of data independence, and web services. HTML, which inspired XML, has been impacted, with XHTML the result. Simply put, XHTML is HTML but following the rules for a well-formed XML document.

An example of XHTML: (303 characters)

```
<html xmlns="http://www.w3.org/1999/xhtml" xml:lang="en" lang="en">
<head>
<meta http-equiv="Content-Type" content="text/html; charset=iso-8859-1"
/>
<title> Frameset DTD XHTML Example </title>
</head>
<frameset cols="100,*">
<frame src="toc.html" />
<frame src="intro.html" name="content" />
</frameset>
</html>
```

the XString representation is: (240 characters)

```
/html@xmlns="http://www.w3.org/1999/xhtml"@xml:lang=en@lang=en

/head/meta@http-equiv=Content-Type@content="text/html; charset=iso-
8859-1"

/title'Frameset DTD XHTML Example
|framset@cols=100,*/frame@src=toc.html|frame@src=intro.html@namecontent
```

The compact representation can be the form the XHTML is stored in on the server. When a client browser requests a web page, the web server can convert the XString representation of the XHTML. Possibly the browser could read the XString and convert it once it's on the browser client, reducing the transfer time. The server could cache frequently accessed XHTML documents as XString in memory buffer or queue, converting them when requested.

### 3.3 Database documents as XML

XML documents in a more general application, can be converted to XString and stored in a relational database as a very long string, often database fields
in a table are defined as "VARCHAR." One other use for XString is to convert
a database table to an XML document, and then to XString. This would allow
for a table of tables.

3.4 Messaging as SOAP/XML-RPC

A very common use for XML (and much hyped) is in inter-platform
communication, for application inter-operability. Two widely used approaches
are the Simple Object Access Protocol (SOAP) [SOAP] and XML remote procedure
call (XML-RPC).

Consider an example of XML-RPC requiring 195 characters

```
<methodCall>
<methodName>sumAndDifference</methodName>
<params>
<param><value><int>5</int></value></param>
<param><value><int>3</int></value></param>
</params>
</methodCall>
```

it compresses in XString to 82 characters as

```
/methodCall/methodName'sumAndDifference|params/param/value/int'5|param/
value/int'3
```

Next, an example SOAP request requiring 481 characters

```
<SOAP-ENV:Envelope
xmlns:SOAP-ENV="http://schemas.xmlsoap.org/soap/envelope/"
SOAP-ENV:encodingStyle="http://schemas.xmlsoap.org/soap/encoding/"/>
<SOAP-ENV:Header>
<t:Transaction
xmlns:t="some-URI"
SOAP-ENV:mustUnderstand="1">
5
</t:Transaction>
</SOAP-ENV:Header>
<SOAP-ENV:Body>
<m:GetLastTradePrice xmlns:m="Some-URI">
<symbol>DEF</symbol>
</m:GetLastTradePrice>
</SOAP-ENV:Body>
</SOAP-ENV:Envelope>
```

compresses in Xstring to only 281 characters as

```
/SOAP-ENV:Envelope@xmlns:SOAP-
ENV="http://schemas.xmlsoap.org/soap/envelope/"
@SOAP-ENV:encodingStyle="http://schemas.xmlsoap.org/soap/encoding/"
/SOAP-ENV:Header/t:Transaction@xmlns:t=some-URI@SOAP-
ENV:mustUnderstand=1'5
|SOAP-ENV:Body/m:GetLastTradePrice xmlns:m=Some-URI/symbol'DEF
```

## 3.5 XML within XML

One interesting application of XString is to embed an XML document within an XML document. This seems to be a very counter-intuitive use of XString, but it allows for XML to represent XML but unambiguously.

### 3.5.1 XML Schema within XML

XML documents can be defined using XML with an XML Schema. The XML Schema is a separate XML document from an instance of a document that follows the XML Schema.

An example XML Schema is: (977 characters)

```
<xs:scheman XMLns:xs="http://www.w3.org/2001/XMLSchema">
  <xs:element name="shiporder">
     <xs:complexType>
       <xs:sequence>
         <xs:element name="orderperson" type="xs:string"/>
         <xs:element name="shipto">
         <xs:complexType>
           <xs:sequence>
             <xs:element name="name" type="xs:string"/>
             <xs:element name="address" type="xs:string"/>
             <xs:element name="city" type="xs:string"/>
             <xs:element name="country" type="xs:string"/>
           </xs:sequence>
         </xs:complexType>
         </xs:element>
         <xs:element name="item" maxOccurs="unbounded">
         <xs:complexType>
         <xs:sequence>
           <xs:element name="title" type="xs:string"/>
           <xs:element name="note" type="xs:string" minOccurs="0"/>
           <xs:element name="quantity" type="xs:positiveInteger"/>
           <xs:element name="price" type="xs:decimal"/>
         </xs:sequence>
         </xs:complexType>
         </xs:element>
       </xs:sequence>
       <xs:attribute name="orderid" type="xs:string" use="required"/>
     </xs:complexType>
  </xs:element>
</xs:schema>
```

as XString the XML Schema is: (665 characters):

```
/xs:schema@xmlns:xs="http://www.w3.org/2001/XMLSchema"
/xs:element@name=shiporder+10/xs:complexType+9/xs:sequence+8
/xs:element@name=orderperson@type=xs:string+0/xs:element@name=shipto+6
/xs:complexType+5/xs:sequence+4
/xs:element@name=name@type=xs:string
|xs:element@name=address@type=xs:string
|xs:element@name=city@type=xs:string
|xs:element@name=country@type=xs:string
/xs:element@name=item@maxOccurs=unbounded+6/xs:complexType+5/xs:sequenc
e+4
/xs:element@name=title@type=xs:string
|xs:element@name=note@type=xs:string@minOccurs=0
|xs:element@name=quantity@type=xs:positiveInteger
|xs:element@name=price@type=xs:decimal
|xs:attribute@name=orderid@type=xs:string@use=required
```

The XML Schema as XString uses depth prefix indicators, and sibling nodes are used when the depth is zero (empty tag, or tags at the same level as siblings).

The XML Schema represents a shipping order. An instance that follows the XML Schema is: (649 characters)

```
<shiporder orderid="889923"
xmlns:xsi="http://www.w3.org/2001/XMLSchema-instance"
xsi:noNamespaceSchemaLocation="shiporder.xsd">
<orderperson>John Smith</orderperson>
<shipto>
<name>Ola Nordmann</name>
<address>Langgt 23</address>
<city>4000 Stavanger</city>
<country>Norway</country>
</shipto>
<item>
<title>Empire Burlesque</title>
<note>Special Edition</note>
<quantity>1</quantity>
<price>10.90</price>
</item>
<item>
<title>Hide your heart</title>
<quantity>1</quantity>
<price>9.90</price>
</item>
</shiporder>
```

as XString: (390 characters)

```
/shiporder@orderid=889923@xmlns:xsi="http://www.w3.org/2001/XMLSchemain
stance"@
xsi:noNamespaceSchemaLocation=shiporder.xsd
/orderperson'John Smith
|shipto+8/name'Ola Nordman|address'Langgt 23|city'4000 S
tavanger
|country'Norway
|item+8/title'Empire Burlesque|note'Special 
Edition|quantity'1|price'10.90
|item+6/title'Hide your heart|quantity'1|price'9.90
```

Within the XString, the spaces have been replaced with the entity   for
a white space.

One problem is that there is a "separation of concerns." Without the XML
Schema for the document, information about the structure of the document is
limited to the specific instance of XML.

Using XString, the instance XML document of the XML Schema can contain the
XML Schema as a XString. An example of XML with the XML Schema embedded is:

```
<shiporder orderid="889923" length="665"
xschema="/xs:schema@xmlns:xs="http://www.w3.org/2001/XMLSchema"
/xs:element@name=shiporder+10/xs:complexType+9/xs:sequence+8
/xs:element@name=orderperson@type=xs:string+0/xs:element
@name=shipto+6/xs:complexType+5/xs:sequence+4/xs:element@name=name
@type=xs:string|xs:element@name=address@type=xs:string|xs:element
@name=city@type=xs:string|xs:element@name=country@type=xs:string
/xs:element@name=item@maxOccurs=unbounded+6/xs:complexType+5
/xs:sequence+4/xs:element@name=title@type=xs:string|xs:element
@name=note@type=xs:string@minOccurs=0|xs:element@name=quantity
@type=xs:positiveInteger|xs:element@name=price@type=xs:decimal
|xs:attribute@name=orderid@type=xs:string@use=required">
<orderperson>John Smith</orderperson>
<shipto>
<name>Ola Nordmann</name>
<address>Langgt 23</address>
<city>4000 Stavanger</city>
<country>Norway</country>
</shipto>
<item>
<title>Empire Burlesque</title>
<note>Special Edition</note>
<quantity>1</quantity>
<price>10.90</price>
</item>
<item>
<title>Hide your heart</title>
<quantity>1</quantity>
<price>9.90</price>
</item>
</shiporder>
```

The revised XML document does not include the XML Schema reference, as the included XML Schema is the "local" XML Schema the document is using. The length of the XString is used for an unambiguous means to know the XString's length.

3.5.2 XML within XML Document Folding

Another use for XString is to embed an XML document within an XML document. Converting an XML document to an XString, and then having it contained as an attribute within a tag is termed "folding" the XML.

For example the initial XML document as XHTML could be:

```
<HTML>
<HEAD>
<TITLE>Example XHTML</TITLE>
</HEAD>
<BODY>
<XSTRING LENGTH="0" TEXT=""/>
<CENTER><H3>Example</H3></CENTER>
</BODY>
</HTML>
```

The simple XHTML has the XSTRING tag embedded. If the XHTML is folded, then the XString is:

```
/HTML/HEAD+2/TITLE'Example
XHTML/BODY/XSTRING+0@LENGTH=0@TEXT=""/CENTER/H3'Example
```

The depth marker + is used to clearly delineate the beginning of a sibling node, and end of a previous child node – to avoid any ambiguity in the XString representation of the document. This XString can be embedded within another XHTML document:

```
<HTML>
<HEAD>
<TITLE>Yet another example</TITLE>
</HEAD>
<BODY>
<XSTRING LENGTH="82"
TEXT="/HTML/HEAD+2/TITLE'Example XHTML/BODY/XSTRING+0@LENGTH=0
@TEXT=''/CENTER/H3'Example"/>
<CENTER><H3>This is yet another example</H3></CENTER>
</BODY>
</HTML>
```

Now the previous XHTML page is now embedded or "folded" into the XHTML page. The space has been replaced with an entity within the text   and other special prefix characters, if ambiguous, could be replaced with the entity equivalent within the XString.

If the XHTML including the XString were to be folded into another XHTML
document, the XString within the XHTML would be:

```
<HTML>
<HEAD>
<TITLE>Going too far perhaps?</TITLE>
</HEAD>
<BODY>
<XSTRING LENGTH="226"
TEXT="/HTML/HEAD+2/TITLE'Yet another example
/BODY/XSTRING+0@LENGTH=82@TEXT="/HTML/HEAD+2/TITLE'Example 
XHTML/BODY/XSTRING+0@LENGTH=0@TEXT=''/CENTER/H3'Example"/CENTER
/H3'This is yet another example"/>
<CENTER><H3>This is way too much!</H3></CENTER>
</BODY>
</HTML>
```

The process of embedding and folding XML documents within another can
continue ad infinitum (theoretically), although the resulting text for the
document might appear as computer gibberish at some point.

An alternative means to embed an XML document within another is to embed each
along with its length, along with an attribute indicating the number of
embedded documents. For example:

```
<HTML>
<HEAD>
<TITLE>Example XHTML</TITLE>
</HEAD>
<BODY>
<XSTRING COUNT="0"/>
<CENTER><H3>Example</H3></CENTER>
</BODY>
</HTML>
```

then folding and embedding within another XML document:

```
<HTML>
<HEAD>
<TITLE>Yet another example</TITLE>
</HEAD>
<BODY>
<XSTRING COUNT="1" LENGTH="82"
TEXT="/HTML/HEAD+2/TITLE'Example XHTML/BODY/XSTRING+0@COUNT=0
/CENTER/H3'Example"/>
<CENTER><H3>This is yet another example</H3></CENTER>
</BODY>
</HTML>
```

The COUNT attribute is consistently among all the folded XML documents, and
for each folded XML document a LENGTH and TEXT attribute is present.

The next XML document containing a folded and embedded document is:

```
<HTML>
<HEAD>
<TITLE>Going too far perhaps?</TITLE>
</HEAD>
<BODY>
<XSTRING COUNT="2" LENGTH="82"
TEXT="/HTML/HEAD+2/TITLE'Example XHTML/BODY/XSTRING+0@COUNT=0
/CENTER/H3'Example" LENGTH="109"
TEXT="/HTML/HEAD/TITLE'Yet another example/BODY/CENTER/H3'Thi
s&nb
sp;is yet another example"/>
<CENTER><H3>This is way too much!</H3></CENTER>
</BODY>
</HTML>
```

The significant difference is that the XString does not embed the prior
XString within the <XSTRING> tag. For each folding and embedding of the XML
document, the previous document is not included within the overall XString.
Each folded XML document is an individual XString, with the particular level
of folding represented as the count of the numbers of XString.

4. Conclusions

XString is an extension of XML, representing a more efficient encoding and
representation of marked up data. The markup and structure of the XML
document are preserved, while represented in a more efficient and flat form.

XString mirrors XPath syntactically somewhat, but is not a substitute for
XPath, which addresses parts of an XML document. XString can reduce the size
of an XML document, while still preserving the information content and its
organization. The representation in XString makes it easier to manipulate,
using basic string operators rather than a parser. XString is XML's own
markup, offering compression and ease of access and processing.

A proof of concept for XString compression and decompression has already been
built and tested, but a full-featured prototype needs to be deployed across
multiple Web sites so that extensive timing experiments can be conducted.

Nonetheless, the XString technology offers substantial hope for improved
performance for Web applications using XML.